\documentclass[conference]{IEEEtran}
\IEEEoverridecommandlockouts
\usepackage{cite}
\usepackage{amsmath,amssymb,amsfonts}
\usepackage{graphicx}
\usepackage{textcomp}
\usepackage{xcolor}
\usepackage{epstopdf}

\usepackage{lineno}
\usepackage{algpseudocode}
\usepackage{algorithm}
\usepackage{subfigure}
\usepackage{booktabs}
\usepackage{multirow}
\usepackage{makecell}
\usepackage[T1]{fontenc}
\usepackage[utf8]{inputenc}
\usepackage{authblk}

\def\BibTeX{{\rm B\kern-.05em{\sc i\kern-.025em b}\kern-.08em
    T\kern-.1667em\lower.7ex\hbox{E}\kern-.125emX}}
\begin{document}

\title{Learning-Based Joint User-AP Association and Resource Allocation in Ultra Dense Network
}
\author[$\dagger$]{Zhipeng Cheng}
\author[$\dagger$*]{Minghui LiWang}
\author[$\dagger$]{Ning Chen}
\author[$\dagger$]{Hongyue Lin}
\author[$\dagger$]{Zhibin Gao}
\author[$\dagger$]{Lianfen Huang}
\affil[$\dagger$]{Dept. of Information and Communication Engineering, Xiamen University, Xiamen, China
}
\affil[*]{Dept. of Electrical and Computer Engineering, University of Western Ontario, London, Canada\authorcr Email:\{chengzhipeng@stu.xmu.edu.cn, mliwang@uwo.ca, lfhuang@xmu.edu.cn\}}

\maketitle

\begin{abstract}
With the advantages of Millimeter wave in wireless communication network, the coverage radius and inter-site distance can be further reduced, the ultra dense network (UDN) becomes the mainstream of future networks. The main challenge faced by UDN is the serious inter-site interference, which needs to be carefully addressed by joint user association and resource allocation methods. In this paper, we propose a multi-agent Q-learning based method to jointly optimize the user association and resource allocation in UDN. The deep Q-network is applied to guarantee the convergence of the proposed method. Simulation results reveal the effectiveness of the proposed method and different performances under different simulation parameters are evaluated.

\end{abstract}


\section{Introduction}
As one of the key technologies of 5G, Ultra-Dense Network is used to densely deploy a large number of small base stations (SBSs) in hotspots to increase capacity and achieve seamless coverage[1]. However, the dense deployment of SBSs complicates the problems of user association, radio resource allocation, interference control and mobility management [2].
In UDN, SBSs are deployed in overlapping manner, users can choose to be associated with several adjacent SBSs via multi-connectivity solutions [3]. The system performance is greatly influenced by the user association patterns. With the increasing deployment density of SBSs, the network topology becomes very complicated. Moreover, a large number of interference sources with very close signal strength bring huge interference to users. This requires a better resource allocation strategy for interference control.

Recently, several research works have devoted to tackle the user association and resource allocation in UDN [4]-[11]. In [4], a novel modularity-based user-centric (MUC) clustering is proposed for resource allocation in UDN to maximize the sum rate per resource block. The basic idea of MUC clustering is to decompose the UDN into several subnetwork by the group structure of users. A energy-efficient (EE) resource allocation strategy in UDN is presented in [5], the EE optimization problem is decomposed into two sub-optimization problems of sub-channel allocation and power allocation. These two problem are solved by a two-stage Stackelberg game with a uniform pricing scheme. The Millimeter Wave (mmWave)-based UDN is considered in [6] and [7]. In [6], the joint user association and resource allocation problem are modeled as a mixed-integer programming problem, which take multiple factors (e.g., load balance, user quality of service, EE and cross-tier interference) into consideration. In [7], the joint user association and resource allocation problem are considered in mmWave self-backhauling UDN, a coalition game based algorithm is proposed to maximize network sum rate. Similarly, a joint power allocation and user association strategy using non-cooperative game theory is developed in [8]. The joint user association and resource allocation problem are considered in [9], [10] and [11]. In [9], a unified non-orthogonal multiple access (NOMA) in UDN is proposed, which focuses on the user association and resource allocation. Two case studies are presented to demonstrate the effectiveness of the framework. Joint optimization of user association and dynamic time division duplexing (TDD) for UDN are studied in [10]. Authors decompose the problem into separate subproblems that can be solved in a distributed manner and prove convergence to the global optimum. The more similar to our work is [11], they propose a novel method for user association and resource allocation based on coordinated multipoint (CoMP). A cell-filtering and location-load based clustering methods are used to reduce network complexity. Then a competition-based resource allocation scheme is proposed based on the results of clustering.

In the light of previous works, we focus on the joint optimization of user-AP association and resource allocation in UDN rather than solve the joint problem in a separate manner. Due to the complexity of this joint problem, we propose a learning-based joint optimization algorithm. The main contributions of this paper can be summarized as follows:

\begin{itemize}
	\item We propose a multi-agent Q-learning based solution to solve the joint problem of user-AP association and resource allocation in UDN.
	
	\item We apply deep Q-network to avoid the curse of dimensionality and accelerate convergence.
	
	\item Demonstrate the effectiveness of the proposed method
	with simulation results compared with other methods.
\end{itemize}

The remainder of the paper is organized as follows: In Section \uppercase\expandafter{\romannumeral2}, we present the system model and the joint problem of user-AP association and resource allocation is formulated. In Section \uppercase\expandafter{\romannumeral3}, the multi-agent Q-learning based solution is proposed. Simulation results are shown in Section \uppercase\expandafter{\romannumeral4} and we conclude the paper in Section \uppercase\expandafter{\romannumeral5}.

\section{System model and problem formulation}
\subsection{Network model}N
We consider a typical UDN composed of $M$ access points (APs) and $N$ users. All APs are identical in terms of coverage radius, antenna gain, pathloss model and maximum transmission power. The set of APs and users are denoted as $\mathcal{M}=\{1,2,\dots,M\}$ and $\mathcal{N}=\{1,2,\dots,N\}$, respectively. All APs can operate on $L$ orthogonal subcarriers, the set of subcarriers is denoted as $\mathcal{L}=\{1,2,\dots,L\}$. Assume that the user can access at least one AP and a maximum of $k$ APs, each AP can serve up to $f$ users. For the simplicity of discussion, we assume that each user can only choose at most one subcarrier from one AP at any time and can only access to its neighboring APs. Thus, we define the candidate AP set for user $i$ as $S_{i}=\left\{j | \mathrm{d}_{i, j} \leq r,i \in \mathcal{N}, j \in \mathcal{M}\right\}$, where $d_{i,j}$ is the distance between user $i$ and AP $j$, $r$ is the coverage radius of the AP. Similarly, we define the candidate user set for AP j as $U_{j}$. The maximum transmission power of the AP is $P_{ap}$.

\begin{figure}[t]
	\centering
	\includegraphics[width=2.5in]{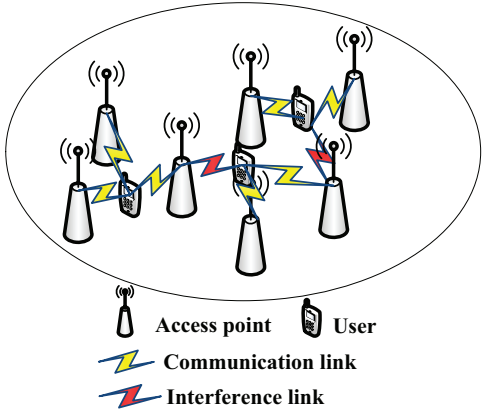}
	\caption{The example layout of UDN.}
	\label{fig_sim}
\end{figure}

 Since all APs are closely deployed, the co-subcarrier interference should be considered, the signal to interference plus noise ratio (SINR) at the user $i$ with AP $j$ using subcarrier $l$ at time $t$ is

\begin{equation}\label{1}
\gamma_{i, j}^{l}(t)=\frac{P_{j} G_{i, j}}{\sum_{j^{\prime} \in M \backslash\{j\}} P_{j^{\prime}} G_{i, j^{\prime}}+N_{0}}
\end{equation}
where $P_{j}=P_{a p} / n_{j}$ is the transmission power of AP $j$, where $n_{j}$ is the number of users associated with AP $j$. $G_{i,j}$ is the channel gain from AP $j$ to user $i$, which incorporates antenna gain, path loss and shadow fading. We consider a flat fading on all subcarriers and thus the channel gains are same from an AP to a user on all subcarriers. $N_{0}$  is the noise power on the subcarrier of bandwidth $W$.

At any time $t$, the transmission capacity of user $i$ with AP $j$ on subcarrier $l$ can be denoted as
\begin{equation}\label{2}
r_{i, j}^{l}(t)=W \log \left(1+\gamma_{i, j}^{l}(t)\right)
\end{equation}

\subsection{Access point selection model}
At any time $t$, all users make AP selection decisions. Here we take a binary indicator $x_{i,j}$   to denote the user-AP association pattern. Let $x_{i,j}=1$ if user $i$ is associated with AP $j$,   otherwise $x_{i,j}=0$ . Note that at any time $t$, we have the following constraints for $x_{i,j}$ as:

\begin{equation}\label{3}
\sum_{j \in S_{i}} x_{i, j} \leq k, \forall i \in \mathcal{N}
\end{equation}

\begin{equation}\label{4}
\sum_{i \in \mathcal{N}} x_{i, j} \leq f, \forall j \in \mathcal{M}
\end{equation}

\begin{equation}\label{5}
x_{i, j} \in\{0,1\}, \forall i \in \mathcal{N}, \forall j \in S_{i}
\end{equation}
where constraint (3) indicates that each user cannot be served more than $k$ APs. Constraint (4) means that one AP can serve simultaneously up to $f$ users.

\subsection{Resource allocation model}
As mentioned above, all subcarriers are shared by the APs, thus the co-subcarrier interference needs to be carefully addressed which greatly limits the system capacity of the UDN. How to effectively allocate $L$ subcarriers to $M$ APs will be the major issue.
At any time $t$, users choose to occupy a subcarrier from an AP. Thus, a binary indicator $y_{i,j}^{l }$  is introduced to indicate the resource allocation strategy. If $y_{i, j}^{l}=1$ means the subcarrier $l$ is allocated to user $i$ from the AP $j$, otherwise $y_{i, j}^{l}=0$. The constraints for $y_{i,j}^{l}$ are as follows:

\begin{equation}\label{6}
\sum_{l \in \mathcal{L}} y_{i, j}^{l} \leq 1, \forall i \in \mathcal{N}, \forall j \in S_{i}
\end{equation}

\begin{equation}\label{7}
y_{i, j}^{l}=1-y_{i^{\prime}, j}^{l}, \forall i, i^{\prime} \in U_{j}, \forall j \in \mathcal{M}
\end{equation}

\begin{equation}\label{8}
\left\{y_{i, j}^{l}=y_{i, j^{\prime}}^{l} | x_{i, j}=x_{i, j^{\prime}}, \forall i \in \mathcal{N}, \forall j, j \in S_{i}\right\}
\end{equation}

\begin{equation}\label{9}
y_{i, j}^{l} \in\{0,1\}, i \in \mathcal{N}, j \in S_{i}, l \in \mathcal{L}
\end{equation}
where constraint (6) states that each user can only choose one subcarrier from its associated AP. Constraint (7) ensures that the user is associated with the same AP use orthogonal subcarriers. Constraint (8) indicates that the APs use the same subcarrier to serve the same user.

Thus, we have the total transmission capacity of the user $i$ at time $t$ as:

\begin{equation}\label{10}
r_{i}(t)=\sum_{j \in S_{i}} \sum_{l \in \mathcal{L}} x_{i, j} y_{i, j}^{l} r_{i, j}^{l}(t), \forall i \in \mathcal{N}
\end{equation}

\subsection{Problem formulation}
The main goal in this paper is to maximize the aggregate network utility while satisfy user's quality of service (QoS) requirements in the UDN. The joint problem of user-AP association and resource allocation is considered. The utility functions are defined for user $i$ at time $t$ as:
\begin{equation}\label{11}
U\left(r_{i}(t)\right)=r_{i}(t)
\end{equation}

 The utility function is a linear function of user $i$'s transmission capacity. As we aim to maximize the long term network utility, we define the long-term reward of user $i$ ${R}_{i}$ as the weighted sum of instantaneous rewards over a finite period $T$:

\begin{equation}\label{12}
R_{i}=\sum_{t=1}^{T} \gamma^{t} U\left(r_{i}(t)\right)
\end{equation}
where $\gamma \in[0,1)$ is the discount factor. Thus, the long-term reward maximization can be formulated as (13).
\begin{equation}\label{13}
\begin{aligned}
	&\max _{x_{i, j}, y_{i, j}^{l}}: \sum_{i=1}^{N} R_{i}\\
	&\text {\textit{ s.t.} }(3)-(9)
\end{aligned}
\end{equation}

Note that due to the non-convex and combinatorial characteristics of the formulated problem, difficulties exist in obtaining a global optimal strategy of this joint problem. In the following section, the multi-agent Q-learning (MAQL) based solution is proposed.

\section{Multi-agent Q-learning based solution}
In this section, we first present the basic idea of MAQL method. Then, the deep Q-network is proposed to avoid the curse of dimensionality.

\subsection{Multi-agent Q-learning method}
In the joint optimization problem of user-AP association and resource allocation, we can model this problem as a discounted stochastic game. In this game, we
assume $N$ users are agents. This $N$ agents stochastic game is defined  by the tuple $\left(S, A_{1}, \dots, A_{N}, r_{1}, \dots, r_{N}, p\right)$. $S$ is the state space, $A_{i}$ and $r_{i}$ are the agent $i$'s action space and reward function. $p$ is the state transition probability. According to the research of stochastic game model, the single-agent Q-learning is extended to a multi-agent scenario, which is called Nash Q-learning. Due to space constraints, refer to [12] for details.

At any time $t$, the user $i$ will observe the current state of the environment and takes action. When all users have taken actions, they observe the reward and the new state, each user then updates its Q table according to

\begin{equation}\label{14}
\begin{aligned}
Q_{i}\left(s, a_{i}\right)=& Q_{i}\left(s, a_{i}\right)+\alpha\left[u_{i}\left(s, a_{i}, \pi_{-i}\right)+\right.\\
&\left.\gamma \max Q_{i}\left(s^{\prime}, a_{i}^{\prime}\right)-Q_{i}\left(s, a_{i}\right)\right]
\end{aligned}
\end{equation}
where $\alpha$ is the learning rate, $u_{i}\left(s, a_{i}, \pi_{-i}\right)$ is the agent $i$'s one-period reward in state $s$ adopting the joint strategies.

As the number of APs and users are fixed in the UDN, if each user obtains the information about reward function and state transition, the Nash equilibrium (NE) can be found to maximize the network utility through message passing within the finite time period $T$ [13]. In the following, we define the agents, states, actions and reward function of our MAQL algorithm.

\begin{itemize}
\item \textbf{Agents:} All $N$ users.
\item \textbf{States:} At time $t$ for user $i$, the state is defined as $s_{i,t}\in \{0,1\}$, indicates whether the user meets its QoS requirement:

		\begin{equation}\label{15}
		s_{i,t}=\left\{\begin{array}{ll}{1,} & {r_{i}(t) \geq r_{Q o S}} \\ {0,} & {r_{i}(t)<r_{Q o S}}\end{array}\right.
		\end{equation}
		
 where $r_{Q o S}$ is the minimum data rate to satisfy the QoS requirement of the user. The number of possible states is $2^{N}$ and this will be very large with large $N$. The state vector can be denoted as $S_{t}=\left\{s_{1, t}, s_{2, t}, \ldots, s_{N, t}\right\}$.

 \item \textbf{Actions:} At time $t$, each user can select up to $k$ APs from the candidate AP set first, then occupy at most one subcarrier from every selected AP. Therefore, the number of possible actions for each user is $M*L$ with one-hot coding. However, the actual action space for AP selection depends on the candidate AP set $S_{i}$.Then, we define the action for user $i$ as

 \begin{equation}\label{16}
a_{i, t}=\left\{m_{i, t}, l_{i, t}\right\}
 \end{equation}

The action vector of $N$ users can be denoted as $A_{t}=\left\{a_{1, t}, a_{2, t}, \ldots, a_{N, t}\right\}$.

\item \textbf{Reward Function: } As we want to maximize the aggregate network utility, then the reward function can be define as $\Psi_{t}=\sum_{i=1}^{N} U\left(r_{i}(t)\right)$.

\end{itemize}

\subsection{Multi-agent deep Q-network for the joint problem}

As can be seen from above, the number of states and actions of the MAQL for the joint problem can be very large for a large $N$, $M$ and $L$. Thus, it is no longer feasible to store the state-action pairs in a Q-table and deep Q-network (DQN) is a better method. The basic idea of DQN is to use the deep neural network (DNN) to represent action and state spaces. DQN takes the advantages of neural network to approximate the action-value function, and uses memory replay to improve the learning performance. Two different neural networks with the same structure, called target-network and evaluated network, are used in DQN. The parameters of the two networks are alternately updated every several steps to improve the learning stability. In each episode, the evaluated Q-network is trained to adapt its parameters to decrease the loss function as follows:

\begin{equation}\label{17}
L_{i}(\theta)=E\left[\left(y_{i}-Q_{i}\left(s, a_{i} ; \theta\right)\right)^{2}\right]
\end{equation}

\begin{equation}\label{18}
y_{i}=u_{i}\left(s, a_{i}\right)+\gamma \max _{a_{i}^{\prime} \in \mathcal{A}_{i}} Q_{i}\left(s^{\prime}, a_{i}^{\prime} ; \theta^{-}\right)
\end{equation}
where $L_{i}(\theta)$ is the loss function, $y_{i}$ is the estimated Q-value of target network. $\theta$, $\theta^{-}$ are the weights of evaluated network and target network, respectively.   

The multi-agent DQN (MADQN) algorithm for the joint problem is summarized in Algorithm 1.

\begin{algorithm}[!htbp]
	\caption{ MADQN  for joint User-AP Association and Resource Allocation}
	\begin{algorithmic}[1]
		\State Initialize learning rate $\alpha$, discount factor $\gamma$, exploration rate $\epsilon$, maximum learning episode $EP$, maximum training steps $T$ per episode.
		\State Initialize the replay memory $D$, evaluated network $Q(s, a ; \theta)$ parameters with random weight $\theta$.
		\State Initialize the target network $Q_{i}\left(s^{\prime}, a_{i}^{\prime} ; \theta^{-}\right)$ with weights $\theta^{-}=\theta$.
		\For {episode=$1:EP$}
		\State Initialize the network state $s$.
		\For{each step=$1:T$}
		\State Each user takes action $a_{i}$ using the $\epsilon$-greedy policy from $Q_{i}(s,a_{i};\theta)$.
		\State Each user obtains the immediate reward $u_{i}$ and new state $s^{\prime}$, and let $s=s^{\prime}$.
		\State Each user stores the transition $(s,a_{i},u_{i},s^{\prime})$ in $D$.
		\State Each user samples random minibatch of transitions $(s,a_{i},u_{i},s^{\prime})$ from $D$.
		\State Each user set $y_{i}$ according to (18).
		\State Each user performs a gradient descent step on $\left(y_{i}-Q_{i}\left(s, a_{i} ; \theta\right)\right)^{2}$ with respect to the network parameters $\theta$.
		\State Every $C$ steps update $\theta^{-} = \theta$
		\EndFor
		\EndFor
	\end{algorithmic}

\end{algorithm}

\section{Performance evaluation}
\subsection{Simulation setup}
In this Section, we conduct the simulation to evaluate the performance of our proposed scheme. We consider a simulation area with a length and width of 50 meters. The APs and users are uniformly distributed within the simulation area.The pathloss model for all APs is $P L=\alpha+10 \beta \log _{10}(d)+\xi[\mathrm{dB}]$
$\xi \sim \mathcal{N}\left(0, \sigma^{2}\right), d$ in meters. $\alpha=72.0, \beta=2.92, \sigma=8.7      \mathrm{dB}$ for NLOS and $\alpha=61.4, \beta=2, \sigma=5.8 \mathrm{dB}$ for LOS [14]. Other important simulation parameters are listed in table \uppercase\expandafter{\romannumeral1}.

\begin{table}[!htbp]
	\caption{Simulation Parameters}
	\centering
	\begin{tabular}{cccc}
		\toprule
		Parameter& Value\\
		\midrule
		Carrier frequency & 28 GHz\\
		Number of subcarriers $L$ & 4 \\
		Subcarrier bandwidth & 180 KHz \\
		Number of APs & 10 \\
		Number of users & 2:2:10\\
		AP radius & 15 m \\
		AP transmission power/Gain & 23 dBm/ 5 dBi\\
		Maximum APs for one user $k$ & 4\\
		Maximum users for one AP $f$ & 4\\
		Noise power density& -174 dBm/Hz\\
		minimum Qos data rate($r_{QoS}$) & 2 Mbps\\
		\bottomrule
	\end{tabular}
\end{table}

The DQN for each learning agent consists of 3 fully connected hidden layers, containing 100, 200, and 50 neurons. The ReLU activation function and RMSProp optimizer are used [15]. We train the learning agent for a total of 400 episodes and 500 steps per episode with a learning rate 0.0001 and a exploration rate from 0.99 to 0.0001. We set the discount factor $\gamma$ to be 0.9.

Due to space constraints, instead of using the state of the art as a comparison, a Max\_RSRP based method is simulated as the baseline to compare the performance with our proposed MADQN method. Users choose to be associated with the $k$ APs with the largest reference signal receiving power (RSRP) and randomly choose a subcarrier in the Max\_RSRP based method. All simulation results are the average result of 50 different user and AP distributions.

\subsection{Simulation results}
\begin{figure}[t]
	\centering
	\includegraphics[width=2.5in]{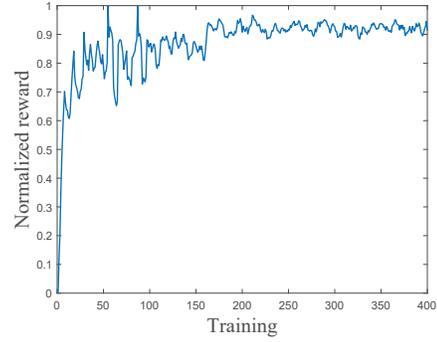}
	\caption{Normalized reward for each episode. The AP number is 5 and the user number is 10.}
	\label{fig_sim}
\end{figure}

Fig. 2 demonstrates the convergence behavior of our proposed MADQN method. As can be seen from the figure, the horizontal axis is the number of training episodes and we take the normalized reward for each episode as the vertical axis. The reward increases with the number of training episodes in the first 150 episodes. when the training episode approximately reaches 200 episodes, the performance gradually converges  despite some fluctuations.

\begin{figure}[htbp]
	\centering
	\subfigure[Total throughput ]{\includegraphics[height=3cm,width=4cm]{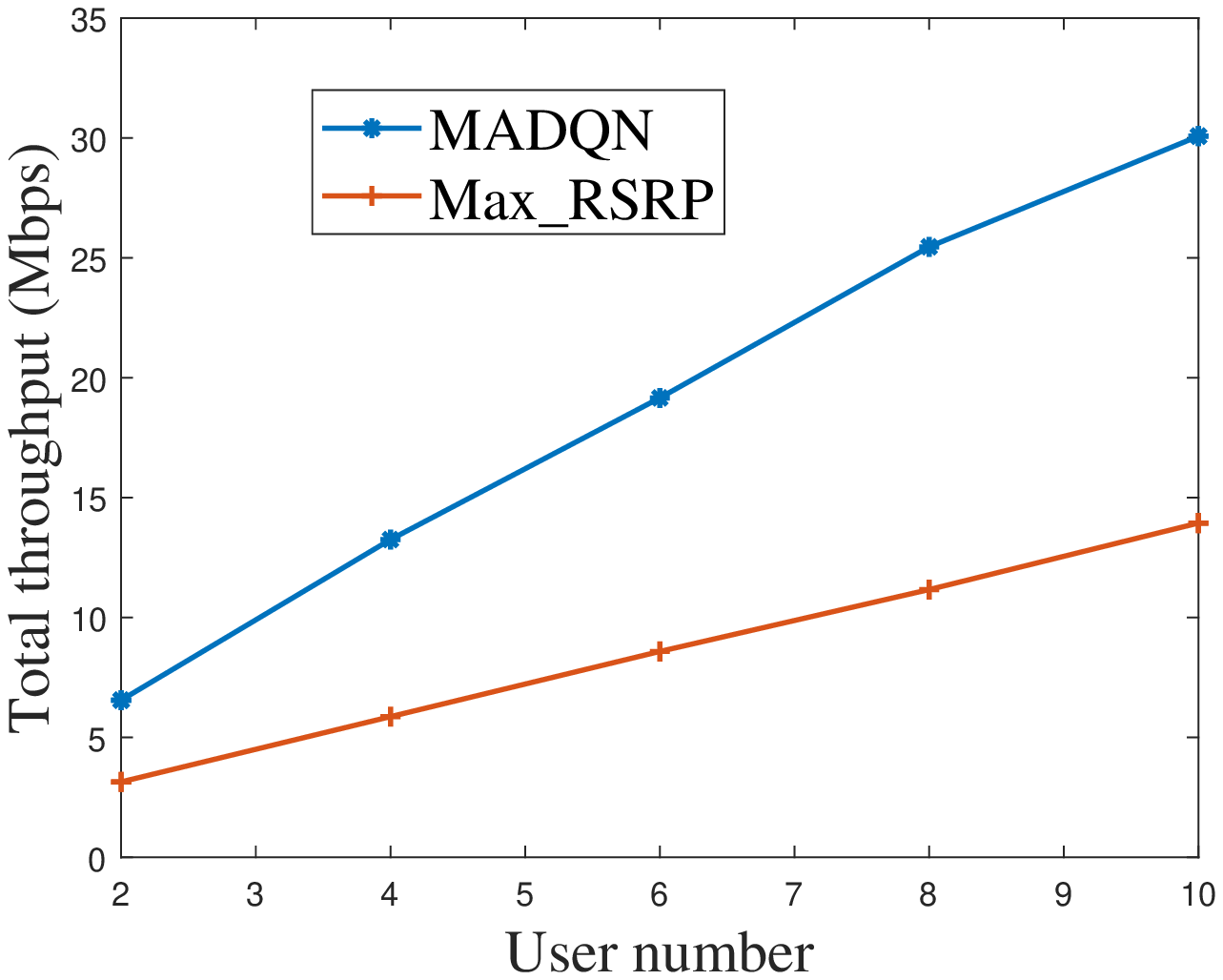}}
	\subfigure[Average user throughput ]{\includegraphics[height=3cm,width=4cm  ]{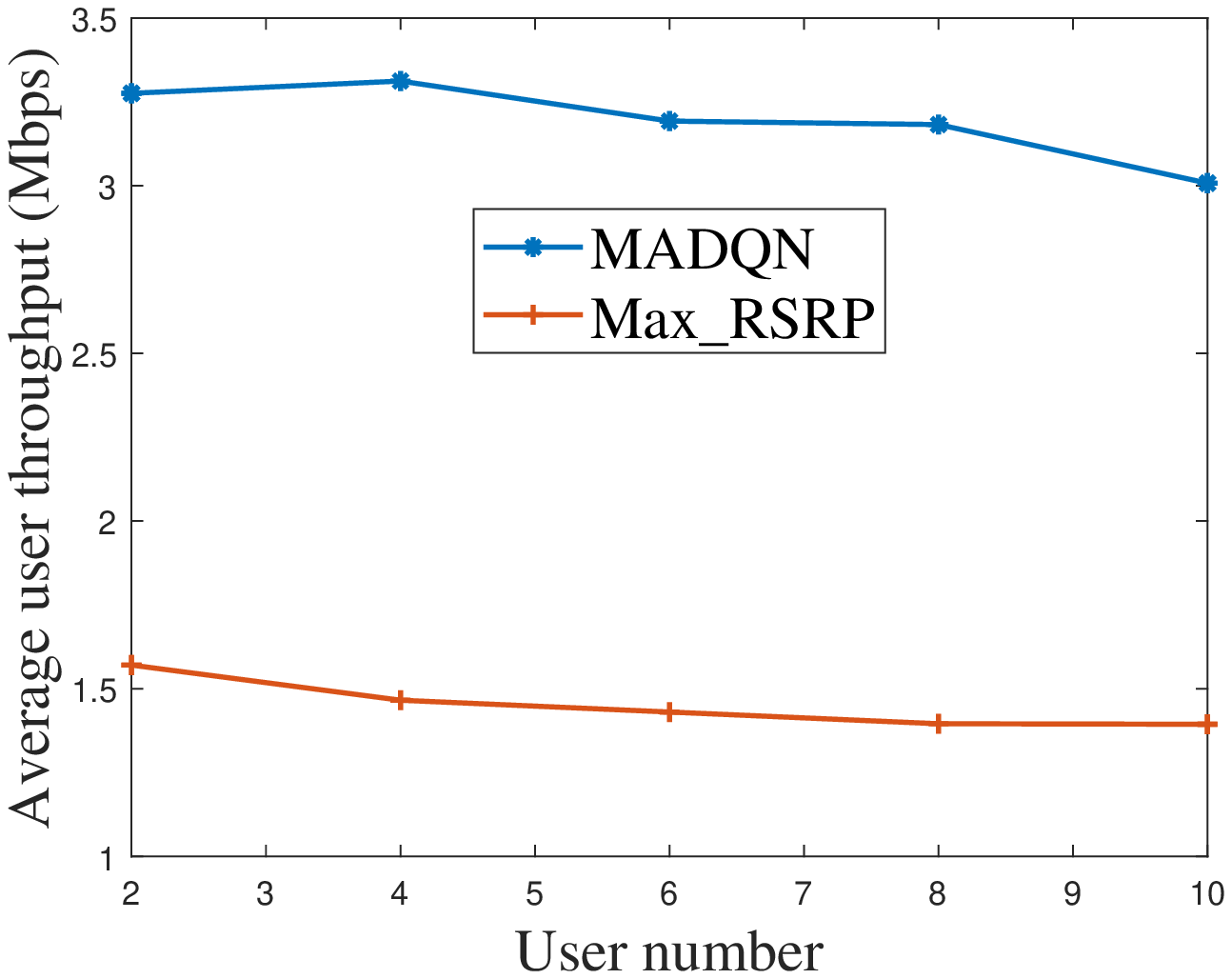}}
	\caption{The throughput versus the user number of two different methods.}
	\label{fig}
\end{figure}

The performances of the two different methods are drawn in Fig. 3. The total throughput and average throughput per user versus the user number of the two methods are compared. As can be seen from Fig. 3, our proposed MADQN method outperforms the Max\_RSRP based method both in total throughput and average user throughput. In particular, the difference between the total throughput is more obvious as the number of users increases. The rationale behind this is that the interference of the Max\_RSRP based method increases rapidly with the number of users.

We further analysis the performance of our proposed MADQN method under different simulation parameters. In Fig. 4, we evaluate the throughput versus the user number for different number of APs. It can be seen that both total throughput and average user throughput increase with the number of APs. The performance of different $k$ and $f$ are evaluated in Fig. 5. The larger $k$ and $f$, the larger the throughput, but this growth is limited as the interference management becomes more complex.

\section{Conclusion}
In this paper, we propose a multi-agent Q-learning (MAQL) based method for the joint problem of user-AP association and resource allocation in ultra dense network. Furthermore, we use the deep Q-network to accelerate the convergence of MAQL. The simulation results demonstrate the feasibility and effectiveness of the proposed method. Moreover, we analysis the performance of our proposed method under different key simulation parameters. This enlightens us to further analyze the relationship between different parameters.

\begin{figure}[t]
	\centering
	\subfigure[Total throughput ]{\includegraphics[height=3cm,width=4cm]{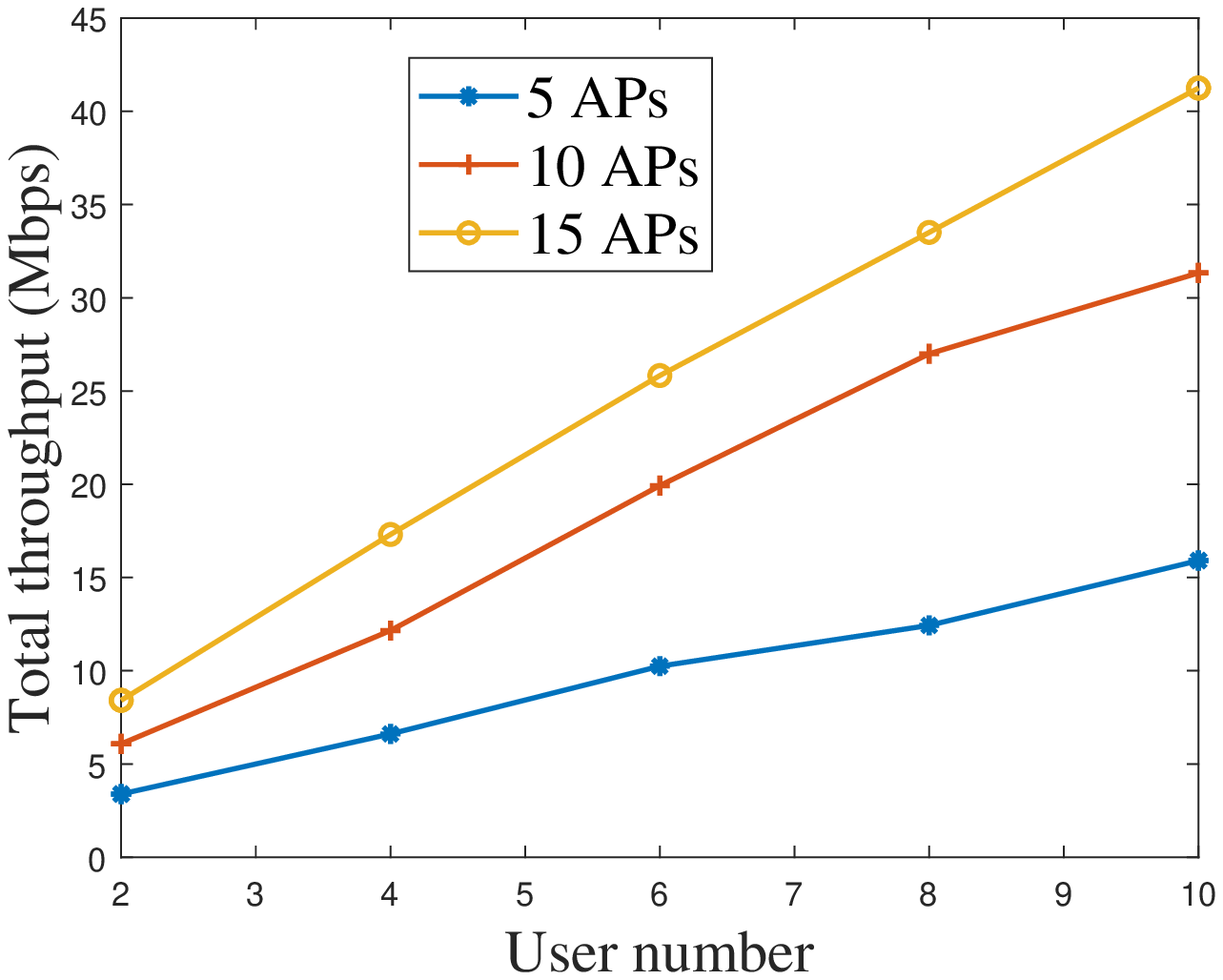}}
	\subfigure[Average user throughput ]{\includegraphics[height=3cm,width=4cm  ]{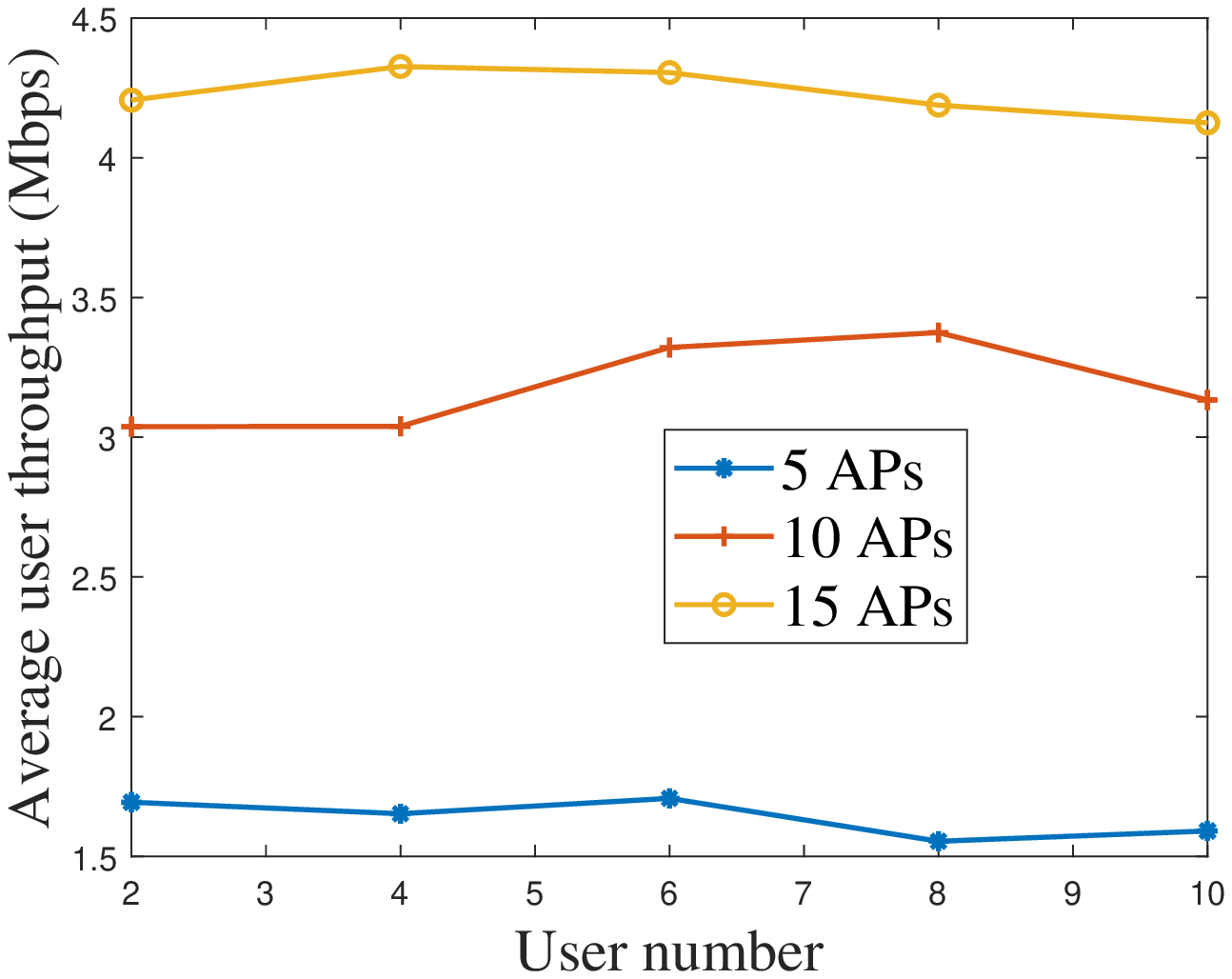}}
	\caption{The throughput versus the user number under different number of APS.}
	\label{fig}
\end{figure}

\begin{figure}[t]
	\centering
	\subfigure[Total throughput ]{\includegraphics[height=3cm,width=4cm]{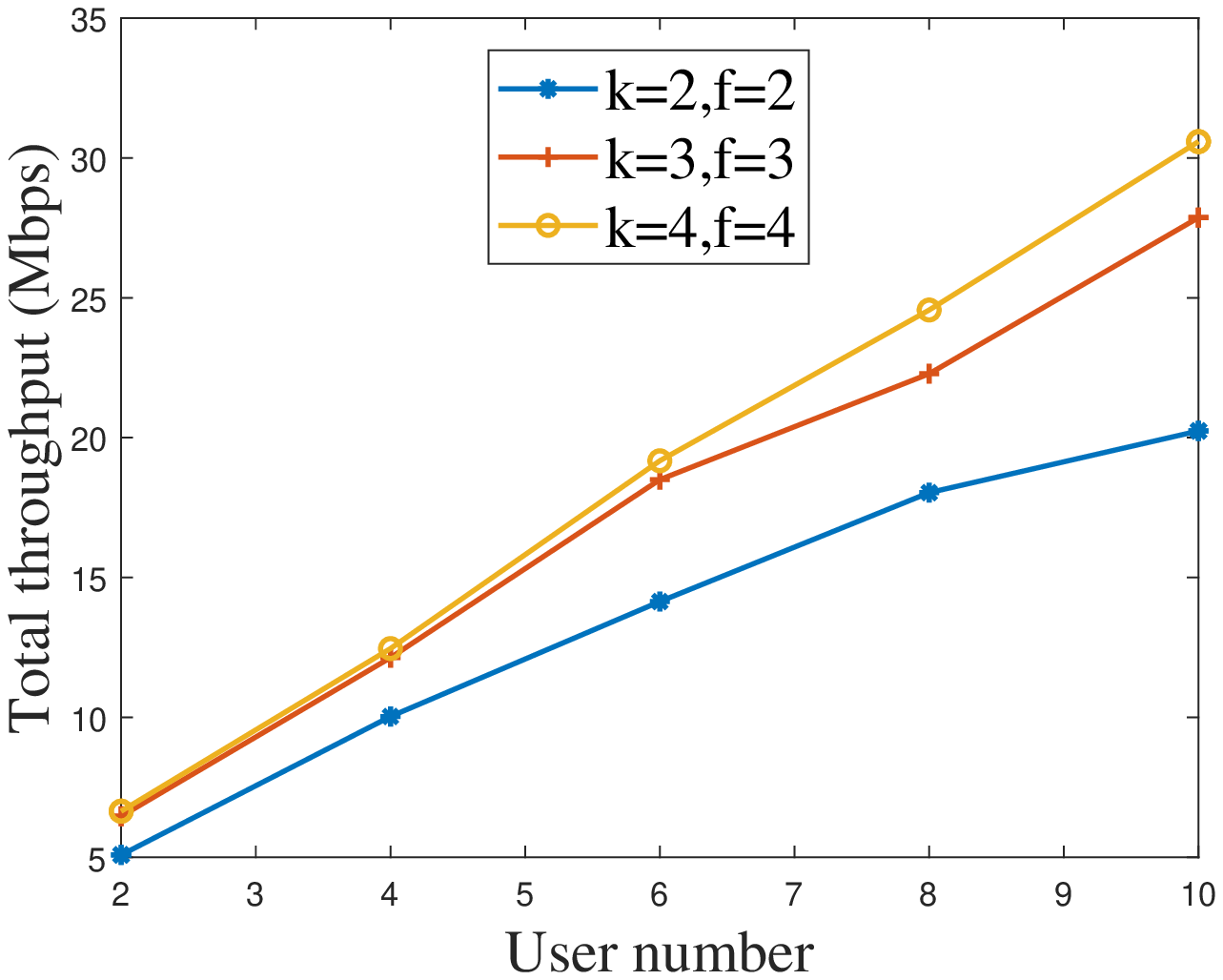}}
	\subfigure[Average user throughput ]{\includegraphics[height=3cm,width=4cm  ]{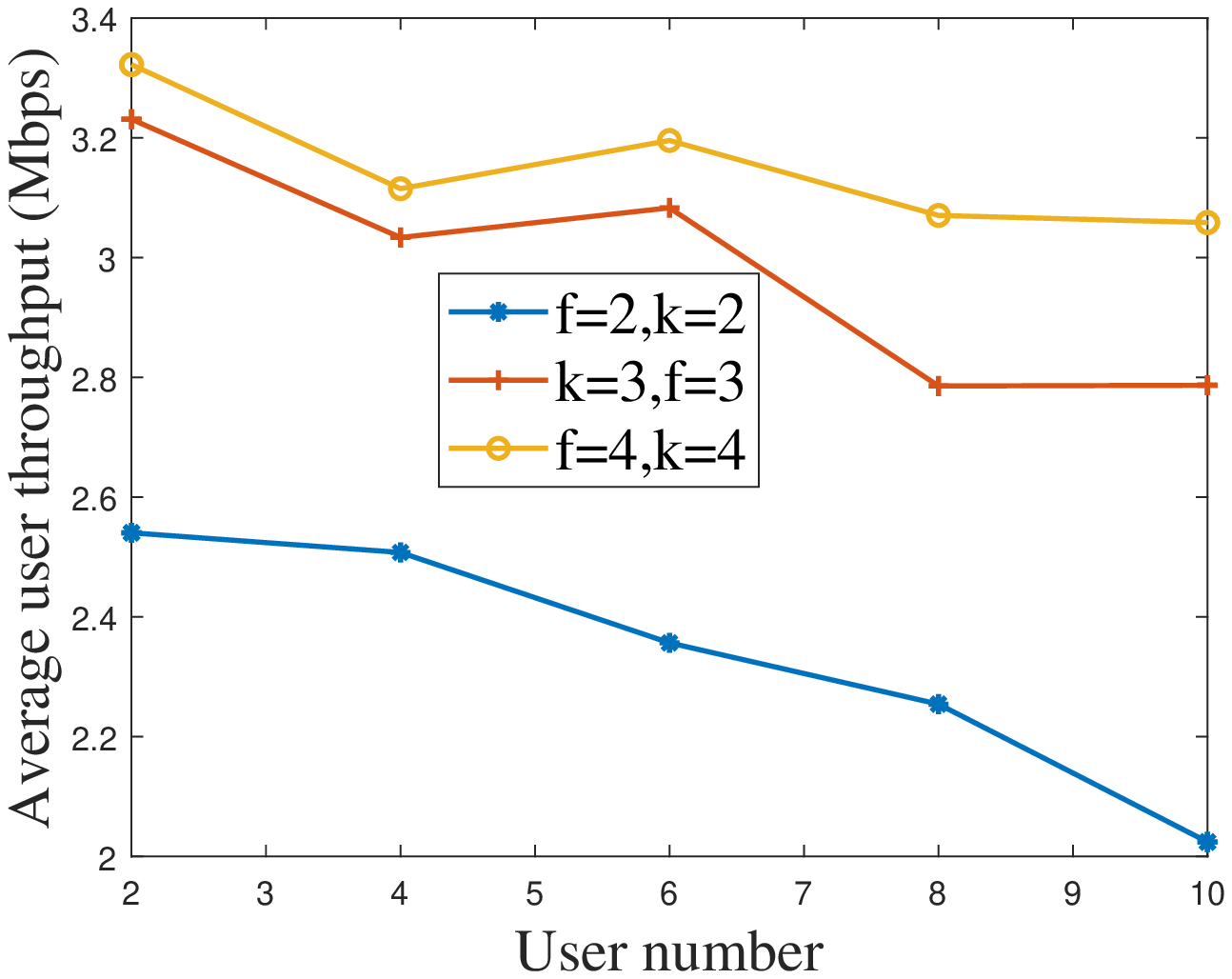}}
	\caption{The throughput versus the user number under different values of $k$ and $f$.}
	\label{fig}
\end{figure}

\section{ACKNOWLEDGMENT}
This work is supported by the National Natural Science Foundation of China under Grant 61971365 and Grant 61871339.


\end{document}